\documentclass[journal=jacsat,manuscript=article]{achemso}

\usepackage[version=3]{mhchem} % Formula subscripts using \ce{}

\usepackage{graphicx}% Include figure files
\usepackage{dcolumn}% Align table columns on decimal point
\usepackage{bm}% bold math
\usepackage[usenames,dvipsnames]{color}
\usepackage{braket}
\usepackage{comment}
\usepackage[english]{babel}
\usepackage{wasysym}
\usepackage[colorlinks,bookmarks=false,citecolor=blue,linkcolor=red,urlcolor=blue]{hyperref}

\usepackage{CJK}

\author{Chong Ye}\email{yechong@bit.edu.cn}
\affiliation{Beijing Key Laboratory of Nanophotonics and Ultrafine Optoelectronic Systems, School of Physics, Beijing Institute of Technology, 100081 Beijing, China}
\author{Xiaowei Mu}
\affiliation{Beijing Key Laboratory of Nanophotonics and Ultrafine Optoelectronic Systems, School of Physics, Beijing Institute of Technology, 100081 Beijing, China}
\author{Yifan Sun}
\affiliation{Beijing Key Laboratory of Nanophotonics and Ultrafine Optoelectronic Systems, School of Physics, Beijing Institute of Technology, 100081 Beijing, China}
\author{Libin Fu}\email{lbfu@gscaep.ac.cn}
\affiliation{Graduate School of China Academy of Engineering Physics, Beijing 100193, China}
\author{Xiangdong Zhang}\email{zhangxd@bit.edu.cn}
\affiliation{Beijing Key Laboratory of Nanophotonics and Ultrafine Optoelectronic Systems, School of Physics, Beijing Institute of Technology, 100081 Beijing, China}

\title[An \textsf{achemso} demo]
  {Enantioselective switch on radiations of dissipative chiral molecules}

\abbreviations{IR,NMR,UV}

\begin{document}

%\begin{tocentry}

%Some journals require a graphical entry for the Table of Contents.
%This should be laid out ``print ready'' so that the sizing of the

%Inside the \texttt{tocentry} environment, the font used is Helvetica
%8\,pt, as required by \emph{Journal of the American Chemical

%The surrounding frame is 9\,cm by 3.5\,cm, which is the maximum
%permitted for  \emph{Journal of the American Chemical Society}
%graphical table of content entries. The box will not resize if the
%content is too big: instead it will overflow the edge of the box.

%This box and the associated title will always be printed on a
%separate page at the end of the document.

%\end{tocentry}

\begin{abstract}
Enantiodetection is an important and challenging task across natural science. Nowadays, some chiroptical methods of enantiodetection based on decoherence-free cyclic three-level models of chiral molecules can reach the ultimate limit of the enantioselectivities in the
molecular responses. They are thus more efficient than traditional chiroptical methods. However, decoherence is inevitable and can severely reduce enantioselectivities in these advanced chiroptical methods, so they only work well in the weak decoherence region. Here, we propose an enantioselective switch on the radiation of dissipative chiral molecules and develop a novel chiroptical method of enantiodetection working well in all decoherence regions. In our scheme, radiation is turned on for the selected enantiomer and simultaneously turned off for its mirror image by designing the electromagnetic fields well based on dissipative cyclic three-level models. The enantiomeric excess of a chiral mixture is determined by comparing its emissions in two cases, where the radiations of two enantiomers are turned off respectively. The corresponding enantioselectivities reach the ultimate limit in all decoherence regions, offering our scheme advantages over other chiroptical methods in enantiodetection. Our work potentially constitutes the starting point for developing more efficient chiroptical techniques for enantiodection in all decoherence regions.
\end{abstract}

\section{Introduction}
Chiral molecules are ubiquitous in nature. They contain two species, named enantiomers, that are mirror images of each other but are not superimposable by translations and rotations. Two enantiomers of the same molecular species share almost identical physical properties, such as melting and boiling points. They play significantly different roles in broad classes of chemical reactions, biological activity, and the function of drugs, making enantiodetection an extremely important and challenging task across natural science~\cite{berova2000circular,barron2009molecular,busch2011chiral,nafie2011vibrational,rhee2009femtosecond,pitzer2013direct,cireasa2015probing,beaulieu2017attosecond,beaulieu2018photoexcitation,arnaboldi2021direct,tang2010optical,tang2011enhanced,wu2020vector,forbes2018optical,forbes2019raman}. 

Nowadays, developing more efficient chiroptical methods, which purely rely on the electric-dipole light-molecule interaction, is becoming an emerging frontier~\cite{ayuso2019synthetic,neufeld2021strong,neufeld2021detecting,khokhlova2022enantiosensitive,ayuso2021ultrafast,ayuso2021enantio,ayuso2022ultrafast,ayuso2022new,jia2011probing,hirota2012triple,patterson2013enantiomer,patterson2013sensitive,patterson2014new,shubert2014identifying,shubert2015rotational,lobsiger2015molecular,shubert2016chiral,chen2020enantio,kang2020effective,ye2021entanglement,chen2022enantiodetection,ye2019determination,cai2022enantiodetection,kral2001cyclic,li2008dynamic,vitanov2019highly,ye2019effective,wu2019robust,wu2020two,torosov2020efficient,torosov2020chiral,liu2022enantiospecific,leibscher2022full,eibenberger2017enantiomer,perez2017coherent,lee2022quantitative,kral2003two,shapiro2000coherently,brumer2001principles,gerbasi2001theory,ye2020fast,ye2021enantio,li2007generalized,li2010theory,liu2021spatial}. 
Some of them~\cite{jia2011probing,hirota2012triple,patterson2013enantiomer,patterson2013sensitive,patterson2014new,shubert2014identifying,shubert2015rotational,lobsiger2015molecular,shubert2016chiral,ye2019determination,chen2020enantio,kang2020effective,ye2021entanglement,chen2022enantiodetection,cai2022enantiodetection,kral2001cyclic,li2008dynamic,vitanov2019highly,ye2019effective,wu2019robust,wu2020two,torosov2020efficient,torosov2020chiral,liu2022enantiospecific,leibscher2022full,eibenberger2017enantiomer,perez2017coherent,lee2022quantitative,kral2003two,shapiro2000coherently,brumer2001principles,gerbasi2001theory,ye2020fast,ye2021enantio,li2007generalized,li2010theory,liu2021spatial} are based on cyclic three-level models of chiral molecules, where the products of the corresponding three electric-dipole transition moments change signs with enantiomers. 
Then, odd- and even-photon processes of chiral molecules therein interfere destructively and constructively for the two enantiomers, making them respond differently to the driving electromagnetic fields. Their strengths can change separately with the strength of each driving electromagnetic field. Therefore, the enantioselectivities,  which measure the relative difference between the responses of two enantiomers, can be freely adjusted and reach the ultimate limit of $100\%$. This offers important advantages over traditional chiroptical methods, such as optical rotation, (vibrational) circular dichroism, and Raman optical activity~\cite{berova2000circular,barron2009molecular,busch2011chiral,nafie2011vibrational}, where the enantioselectivities are tiny and cannot be freely adjusted.  

In the advanced chiroptical methods~\cite{ye2019determination,cai2022enantiodetection,kral2001cyclic,li2008dynamic,vitanov2019highly,ye2019effective,wu2019robust,wu2020two,torosov2020efficient,torosov2020chiral,liu2022enantiospecific,leibscher2022full,kral2003two,shapiro2000coherently,brumer2001principles,gerbasi2001theory,ye2020fast,ye2021enantio,eibenberger2017enantiomer,perez2017coherent,lee2022quantitative}, the ultimate limits of the enantioselectivities are obtained in the framework of the decoherence-free cyclic three-level models. In practice, decoherence is inevitable and can severely reduce enantioselectivities, which is one of the main problems in enantiodetection. To solve this problem, in realistic situations~\cite{patterson2013enantiomer,patterson2013sensitive,patterson2014new,shubert2014identifying,shubert2015rotational,lobsiger2015molecular,shubert2016chiral,eibenberger2017enantiomer,perez2017coherent,lee2022quantitative}, the samples were detected in the weak decoherence region, where the decoherence rates are typically weaker than the coupling strengths. Recently, some of us~\cite{ye2021entanglement} also proposed using frequency-entangled photon pairs as probes to enhance the coincidence signals' enantioselectivities in the strong decoherence region, where the decoherence rates are stronger than the coupling strengths. However, the ultimate limit of enantioselectivity was not reached therein~\cite{ye2021entanglement}. 

In this Letter, we propose a novel chiroptical method of enantiodetection, where the ultimate limit of enantioselectivity is obtained in all decoherence regions. By designing the applied electromagnetic fields well based on the dissipative cyclic three-level models of chiral molecules, we can obtain an enantioselective switch on radiation, i.e., the ultimate limit of enantioselectivity in molecular radiation. In such a case, the radiation is turned on for the selected enantiomer and simultaneously turned off for its mirror image. Because the closed-loop scheme is phase sensitive, the properties of the two enantiomers exchange with each other via increasing the overall phase by $180^{\circ}$. Then, the enantiomeric excess of the chiral mixture can be determined by comparing amplitudes of the radiations in two cases, where the radiations of two enantiomers are turned off respectively. Our numerical simulations show the enantioselective switch on radiation can happen in all decoherence regions, offering our method advantages over other chiroptical methods in enantiodetection.

\section{Enantioselective switch on radiation}
\begin{figure}[htp]
	\centering
	\includegraphics[width=0.5\columnwidth]{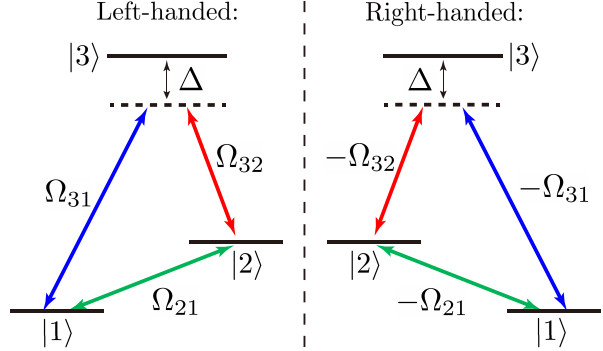}\\
	\caption{Model of the left- and right-handed chiral molecules as cyclic three-level systems. Three electromagnetic fields couple to the electric-dipole
transitions of two enantiomers in a cyclic manner with different coupling strengths $\Omega^{L}_{lj}=-\Omega^{R}_{lj}=\Omega_{lj}$ ($3\ge l>j\ge1$).  }\label{Fig0}
\end{figure}

We consider the enantioselective cyclic three-level models of chiral molecules as shown in Fig.\,\ref{Fig0}, which universally exist in gas-phase asymmetric-top chiral molecules~\cite{patterson2013enantiomer,patterson2013sensitive,patterson2014new,shubert2014identifying,shubert2015rotational,lobsiger2015molecular,shubert2016chiral}. Here, we choose the working states as $|1\rangle=|v_1\rangle|0_{0,0}\rangle$, $|2\rangle=|v_2\rangle|1_{0,0}\rangle$, and $|3\rangle=|v_2\rangle(|1_{1,1}\rangle+|1_{1,-1}\rangle)/\sqrt{2}$~\cite{ye2018real}. The kets $|v_j\rangle$ ($j=1,2,3$) are vibrational sub-levels in the electronic ground state. The kets $|J_{\tau,M}\rangle$ are the rotational eigenfunctions of asymmetric-top molecules, where $J$ is angular momentum quantum number, $M$ is the magnetic quantum number, and $\tau$ running from $-J$ to $J$ in unit steps in the order of increasing energy. By denoting the energies of the working states as $\hbar v_j$, we give the bare transition angular frequency of the transition $|l\rangle\leftrightarrow|j\rangle$ as $v_{lj}=v_l-v_j$ ($l>j$). 

Three electromagnetic fields $\bm{E}_{21}=\bm{e}_z\mathcal{E}_{21}e^{i\omega_{21} t}+c.c.$, $\bm{E}_{31}=\bm{e}_y\mathcal{E}_{31}e^{i\omega_{31} t}+c.c.$, and $\bm{E}_{32}=\bm{e}_x\mathcal{E}_{32}e^{i \omega_{32} t}+c.c.$ with the complex amplitude $\mathcal{E}_{lj}$ are constantly applied to the sample of chiral molecules in the three-photon resonance condition (i.e., $\omega_{31}=\omega_{21}+\omega_{32}$). 
They are near-resonantly coupled with the corresponding transitions in a cyclic manner $|2\rangle\leftrightarrow|1\rangle\leftrightarrow|3\rangle\leftrightarrow|2\rangle$. The other selection-rule-allowed transitions with different bare transition frequencies are off-resonantly coupled with the applied fields and thus are negligible. The polarization directions of the applied fields are well-designed according to the magnetic quantum numbers of the working states~\cite{ye2018real}. The magnetic degenerated states of the working states are thus not coupled to the working models. In this sense, the responses of chiral molecules initially in the Hilbert space of the three working states are governed by the single-loop cyclic three-level models as shown in Fig.\,\ref{Fig0}.

For simplicity and without loss of generality, we choose the angular frequencies of the applied electromagnetic fields as $\omega_{21}=v_{21}$ and $\omega_{31}-\omega_{32}=v_{21}$. In the rotation-wave approximation, the Hamiltonians of the two enantiomers in the interaction picture with respect to $H_{0}=\sum^{3}_{j=1}\hbar v_j |j\rangle\langle j|$ are 
\begin{align}\label{Eq1}
H_{Q}=\hbar\Delta|3\rangle\langle 3|+\sum^{3}_{l>j=1}\hbar\Omega^{Q}_{lj}|l\rangle\langle j|+h.c.,
\end{align}
where $Q=(L,R)$ denotes the molecular chirality. The coupling strengths are given by $\Omega^{Q}_{lj}=\bm{d}^{Q}_{lj}\cdot\bm{\mathcal{E}}_{lj}/(2\hbar)$. The transition electric-dipole moments $\bm{d}^{Q}_{lj}$ change sign with enantiomers (i.e., $\bm{d}^{L}_{lj}=-\bm{d}^{R}_{lj}=\bm{d}_{lj}$). The coupling strengths are thus enantioselective  
\begin{align}\label{Eq2}
\Omega^{L}_{lj}=-\Omega^{R}_{lj}=\Omega_{lj}.
\end{align}
%This means that the overall phases of the cyclic three-level models, which are defined as the phase of the product of the three coupling strengths $\Omega^{Q}_{21}\Omega^{Q}_{32}\Omega^{Q}_{31}$, are different by $180^{\circ}$ for the two enantiomers.

The steady-state equation of a molecule without regard to its chirality is given as 
\begin{align}\label{ESN}
0=-i[H,\rho]+\mathcal{L}(\rho).
\end{align}
Here, $\mathcal{L}(\rho)$ describes the effect of decoherence, which are given as $[\mathcal{L}(\rho)]_{lj}=-(\gamma_{lj}+\gamma^{\mathrm{ph}}_{lj})\rho_{lj}$ and $[\mathcal{L}(\rho)]_{ll}=\sum^{3}_{j^{\prime}=j+1}\Gamma_{jj^{\prime}}\rho_{j^{\prime}j^{\prime}}-\sum^{j-1}_{j^{\prime}=1}\Gamma_{j^{\prime}j}\rho_{jj}$ with $\gamma_{lj}=(\gamma_{l}+\gamma_{j})/2$ and $\gamma_{j}=\sum^{j-1}_{j^{\prime}=1}\Gamma_{j^{\prime}j}$. The coefficient $\Gamma_{lj}$ are pure relaxation rates concerning spontaneous emissions and inelastic collisions. The coefficients $\gamma^{\mathrm{ph}}_{lj}$ are pure dephasing rates attributing to elastic collisions. 

Because the two enantiomers feel different Hamiltonians [see Eq.\,(\ref{Eq1}) and Eq.\,(\ref{Eq2})], they have different steady states. We focus on $\rho_{21}$ in the steady state. It governs the radiation of a single steady-state molecule concerning the induced polarization $\bm{\mathcal{P}}(\omega_{21})=\bm{e}_z{\mathcal{P}}_z(\omega_{21})\propto \rho_{21}$. 
By further designing the applied electromagnetic fields, we expect to obtain that $\rho_{21}$ is $zero$ for one enantiomer and $nonzero$ for its mirror image. In this case, the corresponding radiation is expected to turn on for one enantiomer and turn off for its mirror image, i.e., an enantioselective switch on radiation. 

\begin{figure}[htp]
	\centering
	\includegraphics[width=0.8\columnwidth]{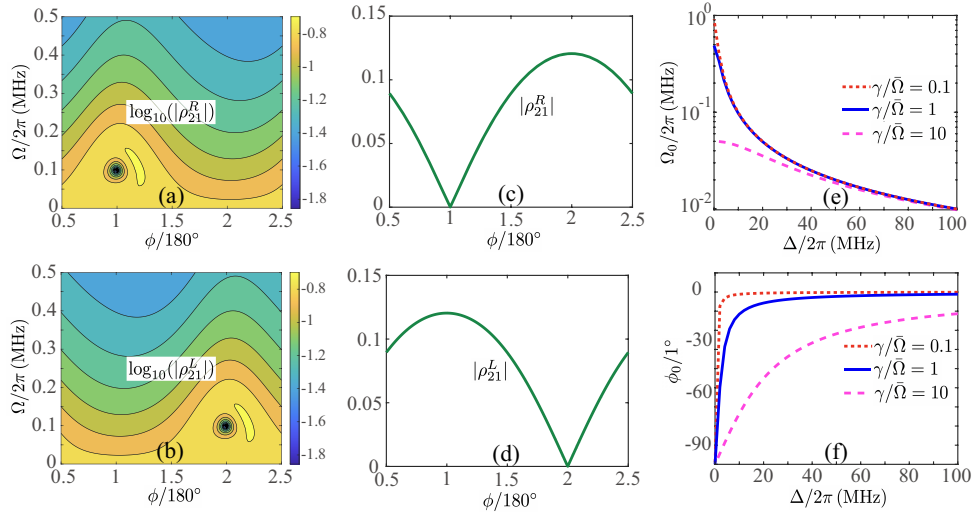}\\
	\caption{$\log_{10}(|\rho_{21}|)$ in the steady state for (a) right- and (b) left-handed molecules in the $\Omega-\phi$ plane with  $\Omega_{21}=\Omega \exp(i\phi)$ ($\Omega\ge0$). For simulations, we take the typical experimental parameters with $\Gamma_{lj}=\gamma^{\mathrm{ph}}_{lj}=\gamma=2\pi\times 0.1$\,MHz, $\Omega_{32}=\Omega_{31}=\bar{\Omega}=2\pi\times 1$\,MHz, and $\Delta=2\pi\times 10$\,MHz. (c) and (d) show the variations of $|\rho_{21}|$ with $\phi$ by fixing $\Omega=\Omega_0\simeq2\pi\times 0.1 $\,MHz. The enantioselective switch on radiation is obtained by tuning the phase to $\phi\simeq 180$ or $\phi\simeq 360^\circ$. (e) and (f) give the amplitude $\Omega_0$ and phase $\phi_0$ as a function of $\Delta$ in the cases, where the left-handed molecules' radiation is turned off in different coherence regions. In (e) and (f), we choose $\Omega_{32}=\Omega_{31}=\bar{\Omega}=2\pi\times 1$\,MHz for simulations.}
	\label{Fig1}
\end{figure}

We demonstrate the enantioselective switch on radiation with an example of 1,2-propanediol. The vibrational sub-levels are chosen as $|v_1\rangle=|v_g\rangle$ and $|v_2\rangle=|v_3\rangle=|v_e\rangle$, where kets $|v_g\rangle$ and $|v_e\rangle$ are vibrational ground and first-excited states concerning the motion of OH-stretch. The corresponding bare transition angular frequencies are $v_{21}=2\pi\times 100.9613$\,THz, $v_{31}=2\pi\times100.9621$\,THz, and $v_{32}=2\pi\times 0.8468$\,GHz~\cite{zhang2020evading}. 
For simplicity and without loss of generality, we assume that $\Gamma_{lj}=\gamma^{\mathrm{ph}}_{lj}=\gamma=2\pi\times 0.1$\,MHz and $\Omega_{32}=\Omega_{31}=\bar{\Omega}=2\pi\times 1$\,MHz by taking the typical experimental values~\cite{patterson2013enantiomer,patterson2013sensitive,patterson2014new,shubert2014identifying,shubert2015rotational,lobsiger2015molecular,shubert2016chiral}. Figure\,\ref{Fig1}\,(a) and Figure\,\ref{Fig1}\,(b) give the numerical results of $\log_{10}(|\rho_{21}|)$ for the two enantiomers by solving Eq.\,(\ref{ESN}) in the $\Omega-\phi$ plane at $\Delta=2\pi\times 10$\,MHz. Here, $\Omega(\ge0)$ and $\phi$ are the amplitude and phase of the coupling strength of transition $|1\rangle\leftrightarrow|2\rangle$ with $\Omega_{21}=\Omega \exp(i\phi)$. The results in Fig.\,\ref{Fig1}\,(a) and Fig.\,\ref{Fig1}\,(b) imply that the enantioselective switch on radiation can be obtained by fixing $\Omega=\Omega_0\simeq 2\pi\times 0.1$\,MHz and tuning the phase $\phi$ via a phase modulator. This is described in detail in Fig.\,\ref{Fig1}\,(c) and Fig.\,\ref{Fig1}\,(d). When the phase is adjusted to $\phi=\phi_0\simeq180^{\circ}$, we have $|\rho^{L}_{21}|\ne0$ and $|\rho^{R}_{21}|=0$ [see the left-half panels of Fig.\,\ref{Fig1}\,(c) and Fig.\,\ref{Fig1}\,(d)]. By increasing the phase to $\phi=\phi_0+180^{\circ}\simeq360^{\circ}$, we have $|\rho^{R}_{21}|\ne0$ and $|\rho^{L}_{21}|=0$ [see the right-half panels of Fig.\,\ref{Fig1}\,(c) and Fig.\,\ref{Fig1}\,(d)]. 

Figure\,\ref{Fig1}\,(e) and Figure\,\ref{Fig1}\,(f) show $\Omega_0$ and $\phi_0$ as functions of $\Delta$ in the cases, where the left-handed molecules' radiation is turned off.  We explore the small-decoherence region with $\gamma/\bar{\Omega}=0.1$, the medium-decoherence region with $\gamma/\bar{\Omega}=1$, and the large-decoherence region with $\gamma/\bar{\Omega}=10$. Other parameters ($\Omega_{32}$ and $\Omega_{31}$) for simulation are the same as in Fig.\,\ref{Fig1}\,(a-d). The results in Fig.\,\ref{Fig1}\,(e) and Fig.\,\ref{Fig1}\,(f)  show that enantioselective radiation can be obtained in all decoherence regions at any detuning by designing the electromagnetic fields well in the framework of the dissipative cyclic three-level models of chiral molecules~(\ref{ESN}). 

The physical mechanism of the enantioselective switch is that the enantioselective interference between odd- and even-photon processes of chiral molecules are freely adjustable in the framework of the dissipative cyclic three-level models. More specifically, we focus on the perturbative region, where the coupling strengths are much smaller than the system's detuning or decoherence rates. The induced polarization of a single steady-state molecule can thus be approximately given as~\cite{boyd2020nonlinear} 
\begin{align}\label{EP}
\mathcal{P}^{Q}_{z}(\omega_{21})\simeq\chi^{(1)}_{zz,Q}\mathcal{E}_{21}+\chi^{(2)}_{zxy,Q}\mathcal{E}_{31}\mathcal{E}^{\ast}_{32},
\end{align}
where the two terms correspond to two- and three-photon processes of chiral molecules. The first- and second-order susceptibilities are~\cite{boyd2020nonlinear}
\begin{align}\label{CHI}
&\chi^{(1)}_{\mu\alpha,Q}=\frac{id^{Q,\mu}_{12}d^{Q,\alpha}_{21}}{\varepsilon_0 \hbar G_{21}}(\rho^{\mathrm{ES}}_{11}-\rho^{\mathrm{ES}}_{22}),
~~~~\chi^{(2)}_{\mu\alpha\beta,Q}=\frac{id^{Q,\mu}_{12}d^{Q,\alpha}_{23}d^{Q,\beta}_{31}}{2\varepsilon_0 \hbar^2 G_{21}}(\frac{\rho^{\mathrm{ES}}_{11}}{\Delta-i G_{31}}
-\frac{\rho^{\mathrm{ES}}_{22}}{\Delta+i G_{32}}),
\end{align}
where $\varepsilon_0$ is the permittivity of vacuum, ${\rho}^{\mathrm{ES}}$ is the density matrix in the field-free equilibrium state, and $G_{lj}=\gamma_{lj}+\gamma^{\mathrm{ph}}_{lj}$
is the transverse relaxation rate of the transition $|l\rangle\leftrightarrow | j\rangle$. Because the electric-dipole moments change signs with enantiomers, the first- and second-order susceptibilities are achiral [$\chi^{(1)}_{\mu\alpha,L}=\chi^{(1)}_{\mu\alpha,R}$] and chiral [$\chi^{(2)}_{\mu\alpha\beta,L}=-\chi^{(2)}_{\mu\alpha\beta,R}$], respectively. The two terms in Eq.\,(\ref{EP}) thus interfere destructively and constructively for the two enantiomers, i.e., enantioselective interference between odd- and even-photon processes of chiral molecules. 
On the other hand, the relative strength between the two terms in Eq.\,(\ref{EP}) can be adjusted freely by tuning the relative strength of the applied electromagnetic fields. When the relative strength between the two terms is adjusted to $one$, the enantioselective switch on radiation is obtained, which presents the ultimate limit of enantioselectivity in radiation concerning $\mathcal{P}^{Q}_{z}(\omega_{21})$. According to Eq.\,(\ref{EP}) and Eq.\,(\ref{CHI}), such an ultimate limit of enantioselectivity is obtained near
\begin{align}\label{ESS}
&\Omega_{21}=\pm
\frac{\Omega_{31}\Omega^{\ast}_{32}}{\rho^{\mathrm{ES}}_{11}-\rho^{\mathrm{ES}}_{22}}(\frac{\rho^{\mathrm{ES}}_{11}}{\Delta-i G_{31}}-\frac{\rho^{\mathrm{ES}}_{22}}{\Delta+i G_{32}}),
\end{align}
which reconfirms the fact that the effect of decoherence is included in our scheme. In our simulation, the two excited states are well separated from the ground state, so the field-free equilibrium state of the three-level system is $\hat{\rho}^{\mathrm{ES}}=|1\rangle\langle 1|$ in a wide temperature range (up to room temperature~\cite{zhang2020evading}). Applying Eq.\,(\ref{ESS}) to the case of Figs.\,\ref{Fig1}\,(a-d), we give $\Omega_{21}\simeq2\pi\times 0.1$\,MHz, agreeing well with the exact values of $(\Omega_0,\phi_0)$ therein. 

Although higher-order processes should be added in Eq.\,(\ref{EP}) in the non-perturbative region, our numerical results in Fig.\,\ref{Fig1}\,(e) and Fig.\,\ref{Fig1}\,(f) show the physical mechanism of our scheme
can still work and yield enantioselective switch on radiation in that region. We note that the susceptibilities can be given in cases that go beyond the few-level models~\cite{boyd2020nonlinear}, and so does the physical mechanism of our scheme. In this sense, enantioselective radiation is a universal phenomenon of dissipative chiral molecules. In contrast, in the traditional chiroptical methods~\cite{berova2000circular,barron2009molecular,busch2011chiral,nafie2011vibrational}, the enantioselectivities are much less than $100\%$ and cannot be freely adjusted. Moreover, our results in Fig.\,\ref{Fig1} show that the ultimate limits of enantioselectivities survive in all decoherence regions. This offers an important advantage over other 
proposals based on the decoherence-free cyclic three-level models~\cite{ye2019determination,cai2022enantiodetection,kral2001cyclic,li2008dynamic,vitanov2019highly,ye2019effective,wu2019robust,wu2020two,torosov2020efficient,torosov2020chiral,liu2022enantiospecific,leibscher2022full,kral2003two,shapiro2000coherently,brumer2001principles,gerbasi2001theory,ye2020fast,ye2021enantio}, where the enantioselectivities decrease with the increase of decoherence.

\section{Determine enantiomeric excess}

\begin{figure}[htp]
	\centering
	\includegraphics[width=0.8\columnwidth]{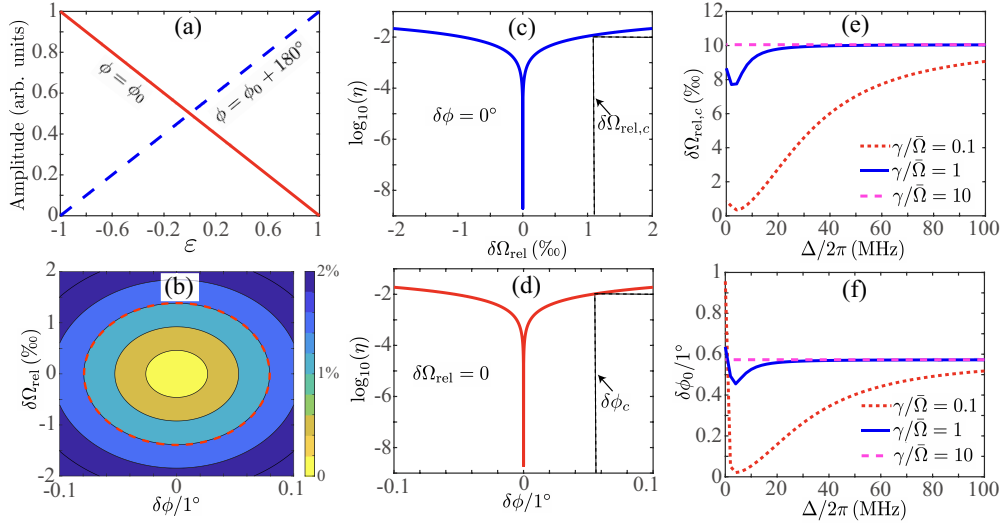}\\
	\caption{(a) Amplitude of the radiation as a function of the enantiomeric excess in two cases, where the radiations of two enantiomers are turned off respectively. (b) Relative error $\eta$ in the $\delta\phi-\delta\Omega_{\mathrm{rel}}$ plane. The red dashed cycle indicates the relative error of $1\%$. (c) $\eta$ as a function of $\delta\Omega_{\mathrm{rel}}$ at $\delta\phi=0^{\circ}$. (d) $\eta$ as a function of $\delta\phi$ at $\delta\Omega_{\mathrm{rel}}=0$. In (c) and (d), we define the critical values of $\delta\Omega_{\mathrm{rel},c}$ and $\delta\phi_{c}$ to achieve the highly efficient estimation of enantiomeric excess ($\eta=1\%$), which characterize the robustness of our scheme. (e) and (f) give $\delta\Omega_{\mathrm{rel},c}$ and $\delta\phi_{c}$ as a function of $\Delta$ at different decoherence regions. }
	\label{Fig2}
\end{figure}

Using the enantioselective switch on radiation, we can determine the enantiomeric excess of a chiral mixture defined as $\varepsilon\equiv(N_L-N_R)/(N_L+N_R)$. Here, $N_L$ and $N_R$ are particle numbers of left- and right-handed molecules in the chiral mixture. For this purpose, we apply two detections on the chiral mixture. In the first detection, the radiation of the left-handed molecule is turned on and that of the right-handed molecule is turned off simultaneously. The amplitude of the radiated electric field at the angular frequency $\omega_{21}$ is $E_{d_1}=N_L {E}^{L}_{0}(\Omega_0,\phi_0)\propto \varepsilon$ [see the blue dashed line in Fig.\,\ref{Fig2}\,(a)]. Here, ${E}^{L}_0(\Omega_0,\phi_0)\ne0$ and ${E}^{R}_0(\Omega_0,\phi_0)=0$ are the emitted signals of single left- and right-handed molecules. In the second detection, the phase is increased by $180^{\circ}$ via a phase modulator. The radiation of the right-handed molecule is turned on with ${E}^{R}_{0}(\Omega_0,\phi_0+180^{\circ})={E}^{L}_{0}(\Omega_0,\phi_0)$ and that of the left-handed molecule is turned off with ${E}^{R}_{0}(\Omega_0,\phi_0+180^{\circ})=0$. In this case, the signal is $E_{d_2}=N_R {E}^{L}_{0}(\Omega_0,\phi_0)\propto-\varepsilon$  [see the red solid line in Fig.\,\ref{Fig2}\,(a)]. 

Therefore, the enantiomeric excess of the sample is given by  
\begin{align}\label{EES}
\varepsilon=\frac{{E}_{d_1}-{E}_{d_2}}{{E}_{d_1}+{E}_{d_2}}.
\end{align}
Our method for determining enantiomeric excess in Eq.\,(\ref{EES}) has three major advantages. Firstly and most importantly, our scheme serves as a sensitive chiroptical method in all decoherence regions because it is based on the signals with an ultimate limit of enantioselectivity in all decoherence regions. Secondly, the enantiomeric excess of the sample is given by applying two detections on itself, i.e.,  without the requirement of hard-to-achieved enantiopure samples as references. In contrast, such references are usually needed in most current chiroptical methods~\cite{berova2000circular,barron2009molecular,busch2011chiral,nafie2011vibrational}. Thirdly, our scheme yields $nonzero$ signals for nearly racemic samples and thus is sensitive to detect their enantiomeric excesses. 
%In contrast, most current chiroptical methods yield weak signals~\cite{hirota2012triple,patterson2013enantiomer,patterson2013sensitive,patterson2014new,shubert2014identifying,shubert2015rotational,lobsiger2015molecular,shubert2016chiral} or suffer large relative errors~\cite{ye2019determination,cai2022enantiodetection} in such samples.

%In contrast, n the advanced schemes~\cite{ye2019determination,cai2022enantiodetection}, the absolute errors reach their maxima at $\varepsilon=0$, yielding them cease to be effective for the nearly racemic samples. This offers our scheme another advantage over the advanced schemes~\cite{ye2019determination,cai2022enantiodetection}. 

%描述怎么\eta, 描述线性。给出高质量的探测	
We also explore the robustness of our scheme against systematic errors, such as the derivations $\delta\phi$ and $\delta\Omega$ from the exact values $(\Omega_{0},\phi_0)$. For this purpose, we define the relative error 
\begin{align}\label{EES1}
\eta\equiv \frac{\delta\varepsilon}{\varepsilon}=2\frac{E^{R}_0(\Omega^{\prime},\phi^{\prime})}{E^{L}_0(\Omega^{\prime},\phi^{\prime})}
\end{align}
with $\Omega^{\prime}=\Omega_0+\delta\Omega$ and $\phi^{\prime}=\phi_0+\delta\phi$. Here, $\delta\varepsilon$ is the error of the estimation resulting from the derivations $\delta\phi$ and $\delta\Omega$. Equation\,(\ref{EES1}) indicates that the relative error is identical for all values of $\varepsilon$. In Fig.\,\ref{Fig2}\,(b), we show the relative errors in the $\delta\phi-\delta\Omega_{\mathrm{rel}}$ plane ($\delta\Omega_{\mathrm{rel}}\equiv\delta\Omega/\Omega_0$). Highly efficient estimation of enantiomeric excess with relative error less than $1\%$ can be obtained in the red dashed cycle. Further, in Fig.\,\ref{Fig2}\,(c), we show $\eta$ as a function of $\delta\Omega_{\mathrm{rel}}$ at $\delta\phi=0^{\circ}$, which shows that the highly efficient estimation is obtained $\delta\Omega_{\mathrm{rel}}\le \delta\Omega_{\mathrm{rel},c}\simeq 1$\textperthousand. In Fig.\,\ref{Fig2}\,(d), we show $\eta$ as a function of $\delta\phi$ at $\delta\Omega_{\mathrm{rel}}=0$, which shows that the highly efficient estimation is obtained $\delta\phi\le \delta\phi_c\simeq 0.5^{\circ}$. We explore the variations of critical values $\delta\Omega_{\mathrm{rel},c}$ [see Fig.\,\ref{Fig2}\,(e)] and $\delta\phi_c$ [see Fig.\,\ref{Fig2}\,(f)] with $\Delta$ in the small-decoherence region $\gamma/\bar{\Omega}=0.1$, the medium-decoherence region $\gamma/\bar{\Omega}=1$, and the large-decoherence region $\gamma/\bar{\Omega}=10$. We find that $\delta\Omega_{\mathrm{rel},c}$ and $\delta\phi_c$ are more sensitive to $\Delta$ in the small-decoherence region. In the large-coherence region, the critical (relative) derivations are almost fixed to $\delta\Omega_{\mathrm{rel},c}\simeq 1\%$ and $\delta\phi_c\simeq0.6^{\circ}$. These results indicate that our scheme should be implemented in the systems in which the precise controls of coupling strengths have been realized~\cite{patterson2013enantiomer,patterson2013sensitive,patterson2014new,shubert2014identifying,shubert2015rotational,lobsiger2015molecular,shubert2016chiral,eibenberger2017enantiomer,perez2017coherent,lee2022quantitative}. 
%描述的不对。

\section{Conclusions}
In summary, we have proposed a scheme to determine the enantiomeric excess of chiral mixtures by using the enantioselective switch on radiation based on the dissipative cyclic three-level model of chiral molecules. The enantioselective switch on radiation is attributed to freely-adjustably enantioselective interference between odd- and even-photon processes of chiral molecules under the illumination of three electromagnetic fields in the three-photon resonance condition. In our scheme, the enantioselectivities reach the ultimate limit of $100\%$, which means the current chiroptical method can be more sensitive than the traditional ones, such as optical rotation, (vibrational) circular dichroism, and Raman optical activity~\cite{berova2000circular,barron2009molecular,busch2011chiral,nafie2011vibrational}. The current ultimate limit of enantioselectivities can survive in all decoherence regions, and so does our scheme of enantiodetection. This offers it an important advantage over other advanced proposals~\cite{ye2019determination,cai2022enantiodetection,kral2001cyclic,li2008dynamic,vitanov2019highly,ye2019effective,wu2019robust,wu2020two,torosov2020efficient,torosov2020chiral,liu2022enantiospecific,leibscher2022full,kral2003two,shapiro2000coherently,brumer2001principles,gerbasi2001theory,ye2020fast,ye2021enantio}, which fail to generate ultimate enantioselectivities in the strong decoherence region. Compared with the recent proposal~\cite{ye2021entanglement} that uses frequency-entangled photon pairs as probes to enhance the coincidence signals’ enantioselectivities, the current scheme has advantages because it uses only the classical source of light and is without the requirement of the coincidence detection. Our work potentially constitutes the starting point for developing more efficient chiroptical techniques for the studies of chiral molecules in all decoherence regions.

\begin{acknowledgement}
This work is supported by the National Natural Science Foundation of China (No.\,91850205, No.\,11725417, and No.\,12088101), the National Science Foundation for Young Scientists of China (No.\,11904022 and No.\,12105011), and Science Challenge Project (No.\,TZ2018005).
\end{acknowledgement}

%\begin{suppinfo}

%\end{suppinfo}

\bibliography{achemso-demo}

\end{document}